\documentclass[prb,twocolumn,showpacs]{revtex4}
\usepackage{amsmath}
\usepackage{graphicx}
\graphicspath{{figures/},{.}}

\begin{document}

\title{Solving rate equations for electron tunneling via discrete
quantum states}
\author{Edgar Bonet}
\author{Mandar M. Deshmukh}
\author{D. C. Ralph}
\affiliation{Laboratory of Atomic and Solid State Physics, Cornell
University, Ithaca, NY, 14853}
\date{\today}

\begin{abstract}
We consider the form of the current-voltage curves generated when
tunneling spectroscopy
is used to measure the energies of individual electronic energy levels
in nanometer-scale systems. We point out that the voltage positions of
the tunneling resonances can undergo temperature-dependent shifts,
leading to errors in spectroscopic measurements that are proportional to
temperature. We do this by solving the set of rate equations that can be
used to describe electron tunneling via discrete quantum states, for a
number of cases important for comparison to experiments, including (1)
when just one spin-degenerate level is accessible for transport, (2)
when 2 spin-degenerate levels are accessible, with no variation in
electron-electron interactions between eigenstates, and (3) when 2
spin-degenerate levels are accessible, but with variations in
electron-electron interactions. We also comment on the general case with
an arbitrary number of accessible levels. In each case we analyze the
voltage-positions, amplitudes, and widths of the current steps due to
the quantum states.
\end{abstract}


\maketitle

\section{Introduction}

Nanometer-scale single-electron tunneling transistors can now be
fabricated in which electron flow occurs through a discrete spectrum of
well-resolved quantum states. This has been achieved in devices
incorporating semiconducting quantum dots, metal nanoparticles, and
molecules.\cite{Ashoori96,Ralph95,Cobden98} In a transistor geometry,
the source-drain voltage $V$ and the gate voltage $V_g$ can be adjusted
to achieve the simplest case that electron flow occurs just through a
single quantum state. As $V$ and $V_g$ are changed, additional excited
electronic states may also become energetically accessible for
tunneling, providing alternative channels for current flow. In this
regime, the tunneling processes can become quite complicated, due to the
many combinations of non-equilibrium states that may be excited during
tunneling, and the possibility of relaxation between these states.

As long as the tunnel-barrier resistances are much greater than $h/e^2$
and internal relaxation is negligible, the currents traveling via any
number of energetically-accessible states can be analyzed in a
sequential-tunneling picture using a rate-equation approach. The general
procedure for completing this type of analysis has been outlined
previously, for example in
Ref.~\onlinecite{Averin91,Beenakker91,Delft01}. Our purpose in this
paper is to present the solutions of this model for selected simple
cases important for analyzing experiments on non-magnetic islands, and
we describe several previously-unappreciated consequences of the model
that explain recent observations. Whenever more than a single
(non-spin-degenerate) quantum state is accessible for tunneling, we show
that the voltage-positions of the tunneling resonances can become
temperature dependent (shifting proportional to $T$). For the important
case of tunneling via one spin-degenerate quantum state, we derive the
full form of the tunneling current as a function of $V$, $V_g$, and
$T$.  This provides simple exact solutions for the voltage
shift, resonance width, and current amplitude, thereby improving upon an
approximate approach used previously. When multiple spin-degenerate
states participate in tunneling, effects of non-equilibrium excitations
and variations in electron-electron interactions can lead to additional
shifts and broadening of the tunneling resonances. The computer code
which we use for calculating the tunneling current in the general case
with an arbitrary number of accessible quantum states is available
electronically in both Mathematica and C formats at
\url{http://www.ccmr.cornell.edu/~ralph/}.

This paper is organized as follows: In section II we review the general
procedure for calculating tunneling currents in the rate-equation
approach. We discuss the physical assumptions under which this approach
is accurate, and we explain our notation. In section III, we solve the
simplest non-trivial case, in which current flow occurs by means of
tunneling via a single spin-degenerate quantum level. In section IV we
then extend this discussion to the case of tunneling via 2 or more
spin-degenerate levels, and we describe several experimentally-relevant
consequences of the rate-equation model for an arbitrary number of
accessible states. In section V, we consider effects of fluctuations in
electron-electron interactions that can occur when current flow
generates non-equilibrium electronic states, and we explain how these
effects can produce additional shifts and can also broaden the measured
tunneling resonances.

\section{Rate-equation calculations of current flow}

\begin{figure}
\centerline{\includegraphics[scale=0.5]{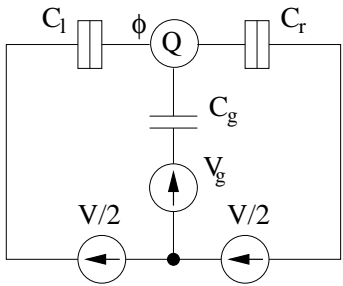}}
\caption{Circuit schematic defining the bias voltage $V$, the gate
voltage $V_g$ and the capacitances $C_l$, $C_r$ and $C_g$. $\phi$ and
$Q$ are the potential and the total charge of the island.}
\label{circuit}
\end{figure}

We are interested in calculating the tunneling current via a non-magnetic
single-electron transistor in the regime where the discrete
quantum states in
the transistor island are well resolved. The circuit under consideration
is shown in Fig.~\ref{circuit}, which illustrates the definitions of the
bias voltage $V$ and the gate voltage $V_g$. We will limit our
discussion to the conditions under which the energy levels are best
resolved: (a) $k_BT$ is smaller than the level spacing, (b) the level
spacing is much smaller than the Coulomb charging energy of the
transistor island $e^2/(2C_\Sigma)$, where $C_\Sigma$ is the total
capacitance of the island, (c) the tunnel barriers
have resistances $\gg h/e^2$ so that cotunneling processes may be
neglected and the tunneling current is accurately described by
lowest-order perturbation theory, and (d) $k_BT$ is larger
than the intrinsic lifetime broadening of the quantum states.
In parts of the discussion, in order
to simplify the notation, we will also assume that electron interactions
are sufficiently weak that many-body eigenstates $| \alpha \rangle$ are
well-approximated as single Slater-determinants specified by the
occupation of a set of single-electron states $i$:
$|\alpha\rangle=\{n_i\}$. Our primary goals are to study the effects on
current flow of non-equilibrium electronic excitations and
electron-electron interactions. Non-equilibrium excitations can be
suppressed when the rate of internal relaxation within the transistor
island is large compared to the tunneling rate. However, measurements on
metal nanoparticles indicate that the relaxation rate is generally
comparable to or slower than the tunneling rate in realistic
samples.\cite{Agam97,Mandar01Al} Therefore we will generally neglect
internal relaxation effects entirely, limiting ourselves to noting the
ways in which internal relaxation will produce qualitative changes to
the results.

\subsection{Energy of the eigenstates}

In general, the quantum-mechanical electronic states
within the transistor island can be complicated correlated many-electron
eigenstates. The energy of any state can be written as a sum of three
terms:
\begin{equation}
E = E_C + E_K + E_J,
\label{E}
\end{equation}
the terms being respectively the electrostatic or ``Coulomb'' energy,
the kinetic energy, and the fluctuations in the
electron-electron interactions. Notice that the mean-field contribution
of the electron-electron interactions is the same as the electrostatic
energy $E_C$. Therefore $E_J$ accounts only for the level-to-level
fluctuations in these interactions.

\subsubsection{Electrostatic energy}

The electrostatic energy will in general depend on the charge of the
island as well as on the applied voltages $V$ and $V_g$. However, what
matters for calculations of electronic transport are energy differences
as electrons make transitions between the transistor island and the
leads. We can select our zero of energy (or, equivalently, the reference
electrostatic potential) for convenience, and we will do so in a way
that makes the energy of the eigenstates on the island independent of
$V$ and $V_g$. The consequence is that the Fermi energies in the leads
will shift with $V$ and $V_g$. To be specific, we choose the reference
electrostatic potential such that $\sum_k C_k \phi_k = 0$, where $C_k$
and $\phi_k$ are the capacitance of the island to the $k$-th lead and
the electric potential of the $k$-th lead, and the sum extends over the
three leads. Using this reference, the charge $Q$ in the island is
related to its potential $\phi$ by $Q = C_\Sigma \phi$. In calculating
the energy required for a tunneling transition, we must consider the
work done. The tunneling of charges $\delta Q_l$ and $\delta Q_r$ from
the island to the left and right leads requires a work
\begin{subequations}
\label{deltaW}
\begin{align}
\delta W &= (\phi_l - \phi) \delta Q_l + (\phi_r - \phi) \delta Q_r \\
		 &= \phi_l \delta Q_l + \phi_r \delta Q_r
			+ \frac{1}{C_\Sigma} Q \delta Q \\
		 &= \delta \left( \phi_l Q_l + \phi_r Q_r
			+ \frac{Q^2}{2 C_\Sigma} \right)
\end{align}
\end{subequations}
where $Q_k$ is the total charge that has tunneled into lead $k$
(Note $\delta Q = - (\delta Q_l + \delta Q_r)$.) From
Eq.~(\ref{deltaW}) it follows that the electrostatic
energy\footnote{Technically, this is an enthalpy, since the work done by
the voltage sources is not taken into account. The real energy of the
field in the capacitors is $\frac{Q^2}{2 C_\Sigma} + \sum_k
\frac{C_k}{2} \phi_k^2$.} of the island is
\begin{equation}
E_C = \frac{Q^2}{2 C_\Sigma}
\end{equation}
and the effective Fermi energies of the leads can be written as
$E^F_k = e \phi_k$, where $e$ is
the electron charge, including its sign. To be explicit,
\begin{subequations}
\label{fermi}
\begin{align}
E^F_l &= +e \frac{2C_r + C_g}{2C_\Sigma} V
         -e \frac{C_g}{C_\Sigma} V_g \\
E^F_r &= -e \frac{2C_l + C_g}{2C_\Sigma} V
         -e \frac{C_g}{C_\Sigma} V_g.
\end{align}
\end{subequations}

Since the charge of the island varies only by multiples of $e$, we can
write it as $Q = Q_0 + Ne$ where $Q_0$ is a background charge and $N$
the number of electrons in the island. The electrostatic energy is then
\begin{equation}
E_C = \frac{1}{2 C_\Sigma} (Q_0 + Ne)^2.
\label{Ec0}
\end{equation}
This is minimized when $N$ is the integer closest to $-Q_0/e$.
Throughout this paper, we will assume that the Coulomb energy is much
larger than the level spacing so that only the two lowest-energy
values
for $N$, namely $N_0$ and $N_1=N_0+1$, are permitted during the
process of current flow.
This assumption allows us to take the electrostatic energy to be proportional
to $N$: since $\big[ N - (N_0 + N_1)/2 \big]^2 = 1/4$ is a constant,
$E_C$ for $N_0$ or $N_1$ electrons can be rewritten, to within a constant, as
\begin{equation}
E_C = N \frac{e}{C_\Sigma} \left( Q_0
		+ \frac{N_0+N_1}{2} e \right)
\label{Ec}
\end{equation}
Notice that Eq.~(\ref{Ec0}) explicitly includes the Coulomb energy which
forbids states not having $N_0$ or $N_1$ electrons, but this is implicit
in Eq.~(\ref{Ec}). The condition $N =
N_0$ or $N_1$ has therefore to be assumed explicitly when using
Eq.~(\ref{Ec}).

\subsubsection{Kinetic energy}

The kinetic energy of the electrons in the island can be written
\begin{equation}
E_K = \sum_i \epsilon_i^K n_i
\end{equation}
where $\epsilon_i^K$ is the energy, relative to the Fermi level, of
spin-degenerate single-electron quantum state $i$, and $n_i$ is the
occupancy of this level (either 0, 1, or 2).

Since $N = \sum_i n_i$, the sum of the electrostatic and kinetic
energies is just
\begin{equation}
E_{CK} = \sum_i \epsilon_i n_i
\label{Eck}
\end{equation}
where $\epsilon_i$ is defined by
\begin{equation}
\epsilon_i = \epsilon_i^K
      + \frac{e}{C_\Sigma} \left( Q_0 +\frac{N_0+N_1}{2} e \right).
\end{equation}
Writing the effective energy of the single-electron states in this
way allows a simple accounting of the average Coulomb energy in the
calculations.

\begin{figure}
\centerline{\includegraphics[scale=0.42]{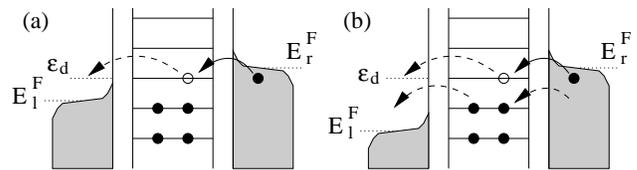}}
\caption{Energy diagrams for the single electron transistor. The island
is represented by a set of discrete energy levels and the leads by
continua of levels. Filled dots in the island stand for electrons
present in an $N_0$-electron ground state. The empty dot is an extra
electron which tunnels onto the island to give an $N_1$-electron state.
The transition marked with a solid arrow is the one which determines
the initial threshold for starting current flow.  The transitions
marked by dotted arrows then also contribute to the total current.
(a) When the Fermi energy of the right lead is swept past the first
level available for tunneling at energy $\epsilon_d$, current can tunnel
through this level. (b) For a slightly lower gate voltage and higher
bias voltage, two levels contribute to tunneling even at the initial
onset of current flow.}
\label{Ediagram}
\end{figure}

In the absence of variations in electron-electron interactions between
electrons in different energy levels, the energy of the island is just
$E_{CK}$. With our conventions, the threshold voltages required for the
onset of a tunneling process can be pictured with simple energy
diagrams, as illustrated in Fig.~\ref{Ediagram}. For example,
at $T=0$, electrons can tunnel
from lead $k$ into the island if the island is a $N_0$-electron state
and Fermi energy $E^F_k$ of lead $k$ is above the energy
$\epsilon_i$ of a non-fully occupied level. In the same way, electrons
can tunnel out of the island into lead $k$ if the island is in a
$N_1$-electron state and $E^F_k$ is below the energy of a
non-empty state.
The onset of the current is associated with the first level available
for tunneling, i.e.\ the lowest-energy non-full level in the $N_0$-electrons
ground state or the highest-energy non-empty level in the
$N_1$-electrons ground
state. As $V$ is ramped for a fixed value of $V_g$, the Fermi energy in a lead
can sweep past the energy required to initiate tunneling via an
eigenstate, producing a stepwise change in current. The
voltage-position, width, and current-amplitude of this step are the
quantities that we will analyze. It is important to note that as $V$
is increased, more than one spin-degenerate quantum level can
contribute to tunneling even at the initial onset of current flow.
One example of this case is illustrated by Fig.~\ref{Ediagram}(b).
The first allowed tunneling transition is for an electron to enter
the level with energy $\epsilon_d$ from the right electrode.
However, after this electron has tunneled in to give a total of $N_1$
electrons on the island, transitions to the left electrode can occur
either from the state with energy $\epsilon_d$ or from the
lower-energy occupied state depicted in Fig.~\ref{Ediagram}(b). If an
electron tunnels out of the lower-energy state, subsequent tunneling
transitions from the right electrode can involve either quantum
level. Therefore, calculations of current for this situation must
include tunneling processes occurring via both levels.

It is possible to have current flow at vanishing $V$ if the Fermi
energy of both leads is aligned with the first level available for
tunneling. The gate voltage that
realizes this condition is called \emph{degeneracy point} and is defined
by
\begin{equation}
V_g^0 = - \frac{C_\Sigma}{e C_g} \epsilon_d
\end{equation}
where $\epsilon_d$ is the energy of this particular level.

\subsubsection{Variations in electron-electron interactions}

In the presence of variations in electron-electron interactions
between electrons in different energy levels,\cite{Agam97} the energy of the
island has the extra term
\begin{equation}
E_J = J(\{n_i\})
\end{equation}
Eq.~(\ref{E}) can be interpreted as an expansion of the energy of the
system around the ground state: the second term is the part of $E$ that
is linear in $\{n_i\}$, the first term is the mean-field contribution of
the quadratic part and $J(\{n_i\})$ is defined to be the rest.
The net effect of the $J(\{n_i\})$ term is to produce shifts in the
energy thresholds for tunneling that depend on the actual state of the
particle. For instance, the effective energy level $\epsilon'_i$
for adding an election to level $i$ starting with the $N_0$-electron
state $\{n_j\}$ is
\begin{equation}
\epsilon'_i = \epsilon_i + J(\{n_j + \delta_{ij}\}) - J(\{n_j\}).
\end{equation}
Notice that this is only defined if $n_i < 2$. In the same way, the
energy of a non-empty energy level in a $N_1$ state can be defined as
\emph{minus} the energy required to remove an electron from that level.

\subsection{Steady-state occupation probabilities}

Because of the influence of the Coulomb charging energy, even in the
simplest cases that we will consider the occupation
probability for a given many-body state $|\alpha\rangle=\{n_i\}$ of the
particle \emph{cannot} be factorized as the product of occupancy
probabilities for each single-electron level. Therefore we have to solve
the full rate-equation problem where the occupation probability of each
many-body state is treated as an independent variable.

The evolution of the occupation probability of state
$| \alpha \rangle$ is given by\cite{Averin91,Beenakker91}
\begin{equation}
\frac{dP_\alpha}{dt} = \sum_\beta
	\left(  \Gamma_{\beta \rightarrow \alpha} P_\beta
	      - \Gamma_{\alpha \rightarrow \beta} P_\alpha \right)
\end{equation}
where $\Gamma_{\alpha \rightarrow \beta}$ is the transition rate from
state $| \alpha \rangle$ to state $| \beta \rangle$.

This can be written in matrix form as
\begin{equation}
\frac{d\mathbf{P}}{dt} = \mathbf{\Gamma} \cdot \mathbf{P}
\label{rate}
\end{equation}
with the following coefficients for the matrix $\mathbf{\Gamma}$:
\begin{subequations}
\begin{align}
\Gamma_{\alpha \beta} &= \Gamma_{\beta \rightarrow \alpha}
	\qquad \text{if} \qquad \alpha \neq \beta \\
\Gamma_{\alpha \alpha} &=
	-\sum_{\beta \neq \alpha} \Gamma_{\alpha \rightarrow \beta}.
\end{align}
\end{subequations}

We do not consider cotunneling or internal relaxation in the particle.
Therefore, the only states that are coupled together are states that
have the same occupancy for all the levels, except one electron
difference in one level. Let's assume that states $|\alpha\rangle$ and
$|\beta\rangle$ differ only by $|\beta\rangle$ having one extra electron
in level $i$. Then
\begin{subequations}
\begin{gather}
\begin{split}
\Gamma_{\alpha \rightarrow \beta}
	& = \gamma^l_i f(\epsilon'_i - E^F_l) (2 - n_i) \\
	& + \gamma^r_i f(\epsilon'_i - E^F_r) (2 - n_i)
	\label{gamma-in}
\end{split} \\
\begin{split}
\Gamma_{\beta \rightarrow \alpha}
	& = \gamma^l_i \big(1 - f(\epsilon'_i - E^F_l)\big) n_i \\
	& + \gamma^r_i \big(1 - f(\epsilon'_i - E^F_r)\big) n_i
	\label{gamma-out}
\end{split}
\end{gather}
\end{subequations}
where
\begin{equation}
f(x)=1/(1+\exp(x/k_BT))
\end{equation}
is the Fermi function corresponding to the temperature in the leads and
$\gamma^l_i$ and $\gamma^r_i$ are the bare tunneling rates between level
$i$ and each of the leads. Here $\epsilon'_i$ is the energy needed to
add an electron to state $|\alpha\rangle$ in level $i$. It includes the
contribution of the interaction term.

The steady-state occupation probabilities can be found by iterating
Eq.~(\ref{rate}) with a
discrete timestep $dt$ to find the probabilities for which $
d\mathbf{P}/dt = 0$. This is equivalent to finding the eigenvector
$\mathbf{P}_0$ of $\mathbf{\Gamma}$ associated with the eigenvalue zero.

\subsection{Current}

Once the occupation probabilities for each state $|\alpha\rangle$ are
determined at given values of $V$ and $V_g$, then the current can be
calculated either through the right tunnel barrier or through
the left barrier. In the steady state these two currents are equal. The
current through the left barrier is\cite{Averin91,Beenakker91}
\begin{equation}
I_l = |e| \sum_\alpha \sum_\beta
	\Gamma^l_{\alpha \rightarrow \beta} P_\alpha
\end{equation}
where $\Gamma^l_{\alpha \rightarrow \beta}$ is the contribution of the
left lead to $\Gamma_{\alpha \rightarrow \beta}$, multiplied by $+1$ or
$-1$ depending on whether the $\alpha \rightarrow \beta$ transition
gives a positive or negative contribution to the current.

In order to get a feeling of the physics that will come out of this
rate-equation model, in the rest of the paper we will consider
selected examples that are simple
enough to be solved by hand, yet have the basic ingredients of
the complete problem.

\section{One spin-degenerate level accessible}

\subsection{General formula}

\begin{figure}
\centerline{\includegraphics[scale=0.5]{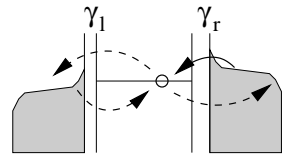}}
\caption{Energy diagram with one level available for tunneling.}
\label{1level}
\end{figure}

Consider the situation represented in Fig.~\ref{1level} where only one
spin-degenerate energy level, with energy $\epsilon_1$,
is accessible for tunneling and (on
account of the large
Coulomb energy) it can be occupied by either zero or one electron, but
not two.\footnote{The case in which one spin-degenerate level is
accessible for tunneling, and the Coulomb energy permits an occupation
of either 1 or 2 electrons (rather than 0 or 1) can be solved by exactly
the same methods:
$I = 2e \frac{\gamma_{r} \gamma_{l}(f_l - f_r)}{ \gamma_{r} (2-f_r) +
\gamma_{l}(2-f_l)}.$}
If we call
\begin{subequations}
\label{deff}
\begin{align}
f_r &= f(\epsilon_1 - E^F_r) \\
f_l &= f(\epsilon_1 - E^F_l)
\end{align}
\end{subequations}
and $N = 0$ or 1 the state with $N$ electrons, the transition rates are
\begin{subequations}
\label{one-lev-gammas}
\begin{align}
\Gamma_{0\rightarrow1} &= 2 \gamma_r f_r + 2 \gamma_l f_l \\
\Gamma_{1\rightarrow0} &= \gamma_r (1 - f_r) + \gamma_l (1 - f_l)
\end{align}
\end{subequations}
for the tunneling-in and tunneling-out transitions.
Then, the occupation probabilities are
\begin{subequations}
\begin{align}
P_1 &= \frac{\Gamma_{0\rightarrow1}}{\Gamma_{0\rightarrow1}
		+ \Gamma_{1\rightarrow0}}
	= \frac{2 \gamma_r f_r + 2 \gamma_l f_l}
		{\gamma_r (1 + f_r) + \gamma_l (1 + f_l)} \\
P_0 &= \frac{\Gamma_{1\rightarrow0}}{\Gamma_{0\rightarrow1}
		+ \Gamma_{1\rightarrow0}}
	= \frac{\gamma_r (1 - f_r) + \gamma_l (1 - f_l)}
		{\gamma_r (1 + f_r) + \gamma_l (1 + f_l)}
\end{align}
\end{subequations}
and the current through the left lead in the steady state is
\begin{align}
I &= |e| \left( \gamma_l (1 - f_l) P_1 - 2 \gamma_l f_l P_0 \right)
	\nonumber \\
   &= 2 |e| \frac{\gamma_r \gamma_l (f_r - f_l)}
		{\gamma_r (1 + f_r) + \gamma_l (1 + f_l)}.
\label{I1level}
\end{align}
This expression differs from an approximate form used in
Ref.~\onlinecite{Deshpande96} to analyze tunneling data.

\begin{figure}
\centerline{\includegraphics[scale=0.7]{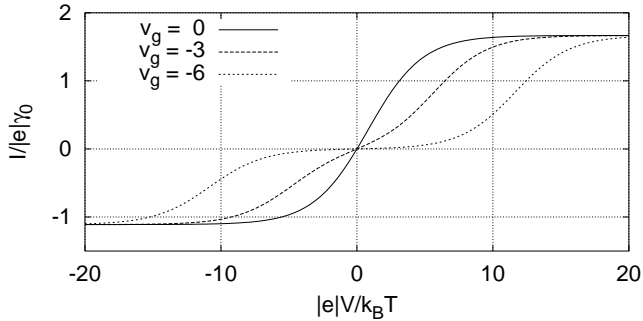}}
\caption{Current profiles as a function of the bias voltage for the case
of a single spin-degenerate level accessible for tunneling, for three
different gate voltages. We assume $C_r=C_l$ and $\gamma_l=4\gamma_r$.
The bias voltage is plotted in units of $k_BT/|e|$. The current is in
units of $|e|\gamma_0$ where $\gamma_0 =
\gamma_l\gamma_r/(\gamma_l+\gamma_r)$. The reduced gate voltage
$v_g=|e|C_g(V_g-V_g^0)/C_\Sigma k_BT$ is 0, -3 or -6.}
\label{steps}
\end{figure}

We can plot the current as a function of the applied voltages by
replacing $f_k$ by their definitions in Eqs.~(\ref{deff}) and
$E^F_k$ by the expressions in Eqs.~(\ref{fermi}). Fig.~\ref{steps}
shows the current steps as a function of the bias voltage when the gate
voltage is first equal to the degeneracy point, then is tuned away from it.

\subsection{High bias limit}

If the level spacing is very large compared to $k_BT$, there is an
interesting regime in which $V$ is substantially bigger than $k_BT/|e|$
yet only one level is involved in the current transport. The limiting
current in
this case is bias-independent and can be obtained from
Eq.~(\ref{I1level}) by setting $f_r=1$ and $f_l=0$ (positive bias) or
$f_l=1$ and $f_r=0$ (negative bias). For these two cases we have
respectively:\cite{Glazman88}
\begin{subequations}
\begin{align}
I_+ &= 2 |e| \frac{\gamma_r \gamma_l}{2 \gamma_r + \gamma_l} \\
I_- &= - 2 |e| \frac{\gamma_r \gamma_l}{\gamma_r + 2 \gamma_l}.
\end{align}
\label{generalform}
\end{subequations}
These expressions give different heights for the positive and negative current
steps. Measuring these heights can therefore allow an experimental
determination of both $\gamma_r$ and $\gamma_l$. Note that this is in
contrast with the case in which tunneling occurs through a single level
that is not spin degenerate. In that case
\begin{equation}
I_1^\pm = \pm |e| \frac {\gamma_r \gamma_l}{\gamma_r + \gamma_l}
\label{Inon-deg}
\end{equation}
for both bias directions,\cite{Glazman88} so that $\gamma_r$ and
$\gamma_l$ can not be determined separately.

In the limit of barriers with very different tunneling rates (which can be
experimentally relevant if the barrier thickness is not well controlled),
the current depends
only on the smaller $\gamma$. For example, if $\gamma_l\gg\gamma_r$,
then $I_+=2|e|\gamma_r$ and $I_-=-|e|\gamma_r$. The factor of 2 in
$I_+/I_-$ arises from the difference in the number of spin states
accessible for tunneling for the rate-limiting transition across the
right barrier.

\subsection{Position and width of the current step}

\begin{figure}
\centerline{\includegraphics[scale=0.5]{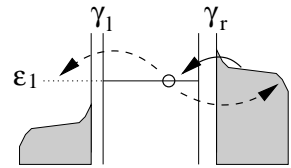}}
\caption{Energy diagram with one level available for tunneling and
$V_g<V_G^0$. Since $E^F_l$ is substantially below $\epsilon_1$,
electrons can tunnel into the island only from the right lead.}
\label{1step}
\end{figure}

\begin{figure}
\centerline{\includegraphics[scale=0.7]{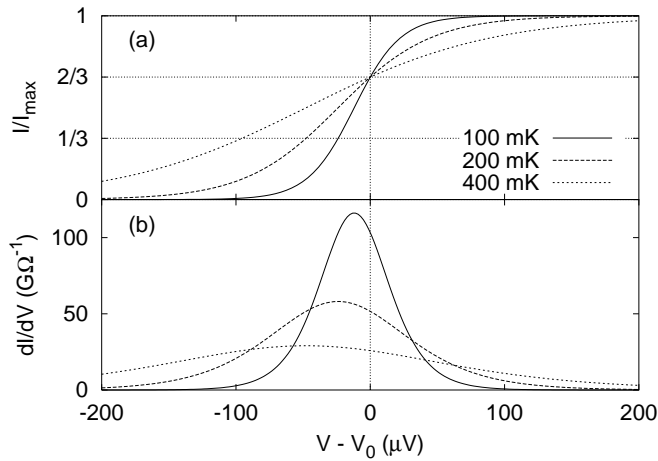}}
\caption{(a) Current step and (b) conductance peak at positive bias and
negative gate voltage for three different temperatures. We assume
$C_r=C_l$, $\gamma_l=50$~MHz\cite{Mandar01Al} and $\gamma_r\gg\gamma_l$. The
peak occurs at $V_0=2C_g(V_g-V_g^0)/C_\Sigma$ at zero temperature and
shifts from this position by an amount $2k_BT\ln2/|e|$ at non-zero
temperature.}
\label{shift}
\end{figure}

Next we consider the case depicted in Fig.~\ref{1step}, in which $V_g$ is
adjusted away from the
degeneracy point so that at
the threshold $V$ for tunneling only the effective Fermi energy in
the right electrode is close to $\epsilon_1$, while the Fermi energy
of the left electrode is at a much lower energy. That is, we will
assume $f_l=0$.
Using this assumption, after some algebra Eq.~(\ref{I1level})
becomes
\begin{equation}
I = I_+ f\left( \epsilon_1 - E^F_r
	- k_B T \ln \frac{2\gamma_r+\gamma_l}{\gamma_r+\gamma_l} \right).
\end{equation}
Even though both spin-states of the quantum level contribute to
tunneling, we can see in this expression that the
current step has the shape of a simple Fermi function whose width is 
given by the
electron temperature of the leads. However, at non-zero temperature, the
center of the step is shifted relative to its position at zero
temperature. The shift is proportional to the temperature, vanishes if
$\gamma_l\gg\gamma_r$, and has a maximum value of $k_BT\ln2$ when
$\gamma_r\gg\gamma_l$. Fig.~\ref{shift} shows the shape of the
conductance peak $dI/dV$ in the latter limit for three different
temperatures.

There is a simple intuitive explanation of the shift in the limit
$\gamma_r\gg\gamma_l$. The current threshold at zero temperature is
given by $E^F_r=\epsilon_1$. At non-zero temperatures, when $E^F_r =
\epsilon_1$, the Fermi occupancy probability is $1/2$ for states in
the right lead with energy $\epsilon_1$. In this case the transition rates
are dominated by electrons tunneling
back and forth from the right lead (since $\gamma_r\gg\gamma_l$):
\begin{subequations}
\begin{align}
\Gamma_{0\rightarrow1} &= \frac{1}{2} (2 \gamma_r) \\
\Gamma_{1\rightarrow0} &= \frac{1}{2} \gamma_r.
\end{align}
\end{subequations}
Here the factor $1/2$ comes from the Fermi occupancy of the lead and
the factor 2 in $\Gamma_{0\rightarrow1}$ is from the spin degeneracy. This
factor is only present for $\Gamma_{0\rightarrow1}$ because electrons
tunneling into the island see two empty states, while an electron
tunneling out comes from
a given spin state. It follows that when $E^F_r=\epsilon_1$ the
probability that the island is in the 1-electron state
is exactly two-thirds. Then the rate at which electrons tunnel
to the left lead (the rate-limiting process determining the total
current) is two-thirds of the maximum value.
This can be seen directly in Fig.~\ref{shift}(a), or in
Fig.~\ref{shift}(b) by the fact that two-thirds of the
current (area under the peaks) lies left of $V-V_0=0$. This
T-dependent shift in the apparent resonance position has been
observed by Deshpande \textit{et al}.\cite{Deshpande00}

\subsection{Zeeman splitting of the energy level}

\begin{figure}
\centerline{\includegraphics[scale=0.7]{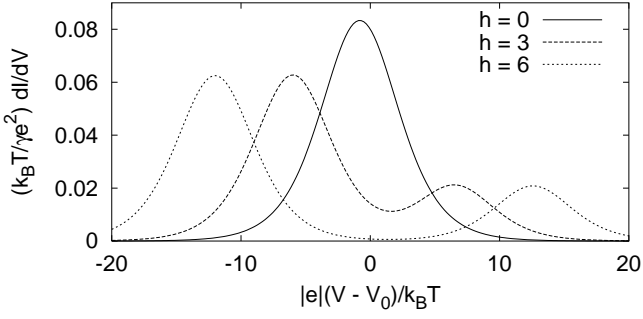}}
\caption{Splitting of a conductance peak in a magnetic field. We assume
$f_l^\pm=0$, $C_r=C_l$, and $\gamma_r=\gamma_l=\gamma$. At zero
temperature and zero field the peak occurs at
$V_0=2C_g(V_g-V_g^0)/C_\Sigma$. The reduced field $h=g\mu_B \mu_0H/(2k_BT)$ is
0, 3 or 6.}
\label{split}
\end{figure}

In the presence of an applied magnetic field, the two spin states
associated with a given orbital level are no longer degenerate, but
split to give the energies $\epsilon^\pm_1 =
\epsilon_1 \pm g\mu_B \mu_0 H/2$. If we call these states $+$
and $-$, and $f^\pm_k \equiv f(\epsilon^\pm_1-E^F_k)$, then the transition
rates are
\begin{subequations}
\begin{align}
\Gamma_{0\rightarrow\pm} &= \gamma_r f^\pm_r + \gamma_l f^\pm_l \\
\Gamma_{\pm\rightarrow0} &= \gamma_r (1-f^\pm_r) + \gamma_l (1-f^\pm_l).
\end{align}
\end{subequations}
Notice the absence of the factors 2 that were in
Eqs.~(\ref{one-lev-gammas}) due to the spin degeneracy. The
occupation probabilities are
\begin{subequations}
\begin{align}
P_0 &= \frac{1}{1
	+\frac{\Gamma_{0\rightarrow+}}{\Gamma_{+\rightarrow0}}
	+\frac{\Gamma_{0\rightarrow-}}{\Gamma_{-\rightarrow0}}} \\
P_\pm &= \frac{\Gamma_{0\rightarrow\pm}}{\Gamma_{\pm\rightarrow0}} P_0
\end{align}
\end{subequations}
and the current through the left lead is
\begin{equation}
I = |e| \gamma_l \big(
	(1-f^+_l) P_+ + (1-f^-_l) P_- - (f^+_l+f^-_l) P_0
	\big)
\end{equation}
Figure~\ref{split} shows the effect of the magnetic field on the
conductance peak at positive bias for a gate voltage below the
degeneracy point (i.e.\ the case $f_l^\pm=0$).
The peak splits into two subpeaks of different weight. This asymmetry can
be understood by noticing that the first subpeak carries a current given
by Eq.~(\ref{Inon-deg}) and the two peaks together give a total
current given by Eq.~(\ref{generalform}). Then the fraction of the
total current carried
by the first subpeak is just
\begin{equation}
\frac{I_1^+}{I_+} = \frac{2\gamma_r+\gamma_l}{2\gamma_r+2\gamma_l}.
\end{equation}
If $\gamma_r\gg\gamma_l$, this ratio is one and the second peak
vanishes.\cite{Glazman88}  On the other hand, if
$\gamma_l\gg\gamma_r$, the peak splits into two subpeaks carrying the same
current.

\section{Two levels accessible}

\begin{figure}
\centerline{\includegraphics[scale=0.5]{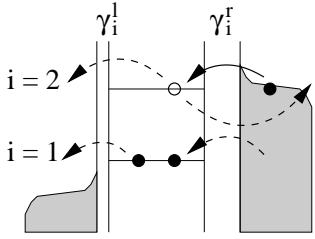}}
\caption{Energy diagram for a case with two levels available for tunneling.}
\label{2levels}
\end{figure}

Next consider the situation pictured in Fig.~\ref{2levels} where two
spin-degenerate levels are accessible for tunneling and the number of
electrons in these levels is $N=2$ or 3.
Because of Coulomb blockade, no current flow is possible until an
electron can tunnel from the right electrode to state 2; however
after this happens both states 1 and 2 can contribute to the current
even at the initial current onset.
Let $(n_1, n_2)$ be the state with $n_1$ electrons in level 1
and $n_2$ electrons in level 2 ($n_1+n_2 = N$), let $P(n_1, n_2)$ be the
probability of state $(n_1, n_2)$, and let $\gamma_i^k$ be the bare tunneling
rate of level $i$ across barrier $k$. We will specialize immediately
to the interesting case of positive bias (as pictured in 
Fig.~\ref{2levels}) with the right
barrier substantially thicker than the left barrier, so $\gamma^l_i
\gg \gamma^r_j$ for $i, j \in \{1, 2\}$. (Note that this is opposite
to the inequality considered in Fig.~\ref{shift}.) To simplify
further, we will also
look only at the current onset at positive bias for a large negative
gate voltage, i.e.\ we will assume
$f(\epsilon_1-E^F_l)=f(\epsilon_2-E^F_l)=0$ and $f(\epsilon_1-E^F_r)=1$.
For this case, $f(\epsilon_2-E^F_r)$ will simply be called $f$. These 
conditions
correspond to line III in the data of Ref.~\onlinecite{Mandar01Al}.

\begin{figure*}
\centerline{\includegraphics{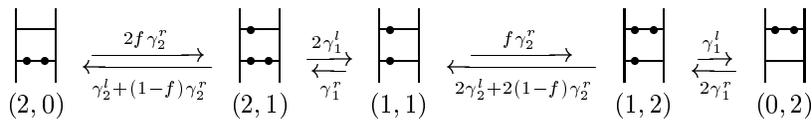}}
\caption{Available transitions for the situation described in
Fig.~\ref{2levels}.}
\label{transitions}
\end{figure*}

Figure~\ref{transitions} shows the available transitions together with the
corresponding transition rates. Since $\gamma^r_i\ll\gamma^l_i$ for
$i=1, 2$, the terms having a factor $1-f$ can be neglected.

\subsection{Rate equation}

The rate equation in this case has to describe eight possible
transitions between five different states. It is therefore
convenient to use the
matrix notation of Eq.~(\ref{rate}), which gives
\begin{subequations}
\label{rate-2levels}
\begin{equation}
\frac{d}{dt}
\begin{pmatrix}
P(2, 0)\\ P(1, 1)\\ P(0, 2)\\ P(2, 1)\\ P(1, 2)
\end{pmatrix}
= \mathbf{\Gamma}
\begin{pmatrix}
P(2, 0)\\ P(1, 1)\\ P(0, 2)\\ P(2, 1)\\ P(1, 2)
\end{pmatrix}
\end{equation}
with
\begin{widetext}
\begin{equation}
\mathbf{\Gamma} =
\left(
\begin{array}{ccc|cc}
-2f\gamma^r_2 & 0 & 0 & \gamma^l_2+(1-f)\gamma^r_2 & 0\\
0 & -\gamma^r_1-f\gamma^r_2 & 0
	& 2\gamma^l_1 & 2\gamma^l_2+2(1-f)\gamma^r_2\\
0 & 0 & -2\gamma^r_1 & 0 & \gamma^l_1\\
\hline
2f\gamma^r_2 & \gamma^r_1 & 0
	& -\gamma^l_2-(1-f)\gamma^r_2-2\gamma^l_1 & 0\\
0 & f\gamma^r_2 & 2\gamma^r_1
	& 0 & -2\gamma^l_2-2(1-f)\gamma^r_2-\gamma^l_1
\end{array}
\right).
\end{equation}
\end{widetext}
\end{subequations}
This matrix has the structure
\begin{equation}
\mathbf{\Gamma} =
\left(
\begin{array}{c|c}
\mathbf{\Gamma}_{uu} & \mathbf{\Gamma}_{uc} \\
\hline
\mathbf{\Gamma}_{cu} & \mathbf{\Gamma}_{cc}
\end{array}
\right)
\label{blocks}
\end{equation}
where $\mathbf{\Gamma}_{uu}$ and $\mathbf{\Gamma}_{cc}$ are diagonal
blocks associated respectively with the $N_0$-electron
(\emph{uncharged}) and $N_1$-electron (\emph{charged}) states. The
cross-diagonal blocks are associated with the tunneling-out
($\mathbf{\Gamma}_{uc}$) and tunneling-in ($\mathbf{\Gamma}_{cu}$)
events. This structure is preserved whatever number of levels are
available for tunneling.

In the steady state, the solutions for the occupation probabilities
are as follows:
\begin{subequations}
\begin{align}
P(2, 1) &\ll 1 &
P(1, 2) &\ll 1 \\
P(2, 0) &= \frac{1}{S} &
P(1, 1) &= \frac{4fK}{S} &
P(0, 2) &= \frac{f^2K^2}{S}
\end{align}
\end{subequations}
where $K = \frac{\gamma^l_1\gamma^r_2}{\gamma^l_2\gamma^r_1}$
and $S = 1 + 4fK + f^2K^2$.

\subsection{Current}

\begin{figure}
\centerline{\includegraphics[scale=0.7]{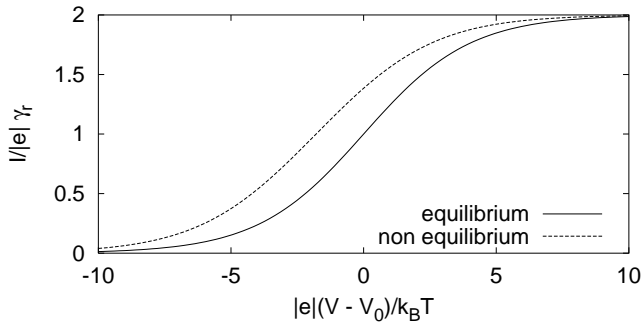}}
\caption{Shift of the current step by non-equilibrium in the 
2-levels-accessible case. We assume
$\gamma_1^l=\gamma_2^l=\gamma_l$ and 
$\gamma_1^r=\gamma_2^r=\gamma_r$, with $\gamma_l\gg\gamma_r$. The step
occurs at $V_0=2C_g(V_g-V_g^0)/C_\Sigma$ at zero temperature. The
``equilibrium'' curve assumes infinitely fast relaxation in the island.
The ``non equilibrium'' curve assumes no relaxation.}
\label{non-eq}
\end{figure}

Since we can neglect the tunneling-out transitions through the right
barrier, we can calculate the current as the sum of the contributions
of the tunneling-in events through this barrier:
\begin{align}
\frac{I}{|e|} &= 2 f \gamma^r_2 P(2, 0)
    + (f \gamma^r_2 + \gamma^r_1) P(1, 1) + 2 \gamma^r_1 P(0, 2)
	\nonumber \\
I &= |e|\frac{(4 \gamma^r_2 K + 2 \gamma^r_1 K^2)f^2
             + (2 \gamma^r_2 + 4 \gamma^r_1)f}
    {1 + 4fK + f^2K^2}.
\label{Inon-eq}
\end{align}
In Fig.~\ref{non-eq} we compare this expression to the current we would
have in the presence of infinitely fast relaxation in the island (state
$(1,1)$ relaxing instantaneously to $(2,0)$). In such a case electrons
can only tunnel into the higher energy level in the island. Since the
tunneling-in of electrons is the rate-limiting process, this situation
is equivalent to the case of Eq.~(\ref{I1level}) when only one level is
accessible for tunneling, and the current would just be
\begin{equation}
I_{\text{equilibrium}} = 2 |e| \gamma^r_2 f(\epsilon_2-E^F_r).
\end{equation}

\noindent
The main effect of non-equilibrium states as illustrated in 
Fig.~\ref{non-eq} is therefore to shift the
current step
to lower voltage. Although not exactly a Fermi function, the shape of the
step described by Eq.~(\ref{Inon-eq}) is very close to a Fermi function,
shifted by $-1.79k_BT$ and widened by
8.5\%. The shift can be understood as follows: When $E^F_r=\epsilon_2$,
electrons tunneling to the upper level come from half-full states in the
right lead. If the island is in a non-equilibrium state ($(1,1)$ or
$(0,2)$), electrons can also tunnel to the lower level. Since these
electrons come from full states in the lead, the current at
$E^F_r=\epsilon_2$ is higher when these states are allowed, hence the
shift.

\begin{figure}
\centerline{\includegraphics[scale=0.7]{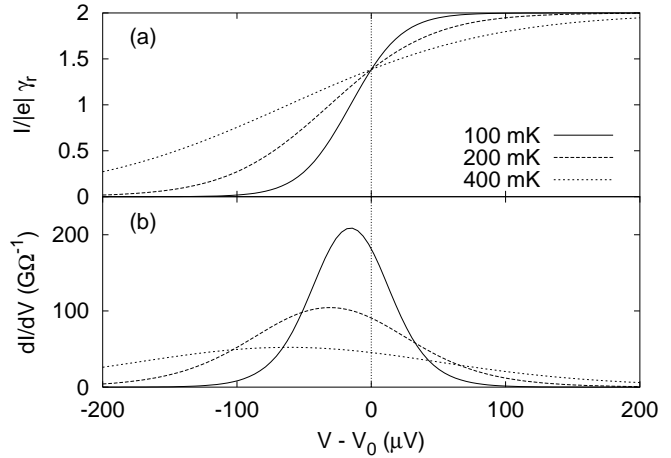}}
\caption{Dependence of (a) the current step and (b) the conductance peak
on the temperature in the 2-levels-accessible case in the presence of 
non-equilibrium. We assume $\gamma_1^l=\gamma_2^l=\gamma_l$ and 
$\gamma_1^r=\gamma_2^r=\gamma_r$, with $\gamma_l\gg\gamma_r$ and no
relaxation in the island.}
\label{non-eq-t}
\end{figure}

The temperature dependence of the current step and the conductance peak
in this two-level-accessible case with $\gamma_l\gg\gamma_r$ is 
displayed in Fig.~\ref{non-eq-t}.
Although the $T$-dependent shift looks very similar to the result for 
one level displayed in
Fig.~\ref{shift}, the shift in Fig.~\ref{non-eq-t}
is of a different nature since it originates from non-equilibrium states.
For the one-level-accessible case, there was no shift for positive 
bias with $\gamma_l\gg\gamma_r$. If we look
at the opposite limit with two levels (positive bias 
$\gamma_l\ll\gamma_r$), the rate equation will be dominated by
electrons tunneling back and forth between the right lead and the second
level in the island. This situation is very similar to the one-level
case and gives the
current:
\begin{equation}
I = |e| (2 \gamma^l_1 + \gamma^l_2)
	f(\epsilon_2 - E^F_r - k_B T \ln 2),
\end{equation}
where the shift by $k_BT\ln2$ is explained by the same argument as in
the one-level case.  The additional level therefore does not produce 
an additional shift when $\gamma_l\ll\gamma_r$.

If the voltages are tuned so that more than two levels are made available for
tunneling-out transitions (by lowering $E^F_l$), or if the tunnel 
couplings to state 1 are greater than to state 2, then the shifting of
the resonance away from the $T=0$ position in the 
$\gamma_l\gg\gamma_r$ case will be enhanced beyond what is shown in 
Fig.~\ref{non-eq-t}.
This shift will however
remain proportional to $k_BT$.

We have also considered the case when $E^F_r$ is very high, so that many
levels are accessible for an electron to tunnel into the island
across the higher-resistance tunnel barrier, while $E^F_l$ remains
fixed slightly below $\epsilon_1$. In such a situation the total
tunneling-in transition rate will be proportional to the number of
levels available for tunneling in, and this rate can eventually
become greater than
the tunneling-out rate which will be roughly constant. In this case, tunneling
though the left lead will eventually become the bottleneck process
even if $\gamma_l \gg \gamma_r$, which allows one to
estimate an average tunneling rate through the lower-resistance barrier even in
the case of very asymmetric barriers.

\section{Two levels accessible with variations in the interactions}

In the presence of variations in electron-electron interactions, the
energy thresholds for tunneling are different depending on whether
the island is initially
in a ground state or in an excited state. For example, in the case
described in the
previous section, this effect can make the energy required for the
$(1,1) \rightarrow (1,2)$ transition different than the $(2,0)
\rightarrow (2,1)$ transition.
We can account for such
variations by assigning a different energy to the upper level in the
presence or absence of an excitation in the island. Namely, the energy
of the upper level will be $\epsilon_2$ for the $(2,0)\rightarrow(2,1)$
transition and $\epsilon'_2=\epsilon_2+\delta$ for the
$(1,1)\rightarrow(1,2)$ transition. Here $\delta$ is a measure of the
strength of the variations. In order to generalize the previous
notation, we will call $f=f(\epsilon_2-E^F_r)$ and
$f'=f(\epsilon'_2-E^F_r)$.

The possible transitions are still described by Fig.~\ref{transitions}
and the corresponding rate equations are the same as
Eqs.~(\ref{rate-2levels}) but with
\begin{subequations}
\begin{align}
\Gamma_{(1,1) \rightarrow (1,2)} &= f' \gamma^r_2 \\
\Gamma_{(1,2) \rightarrow (1,1)} &= 2 \gamma^l_2 + 2(1-f')\gamma^r_2
\end{align}
\end{subequations}
which gives the current
\begin{equation}
I = |e| \frac{(4 \gamma^r_2 K + 2 \gamma^r_1 K^2)ff'
             + (2 \gamma^r_2 + 4 \gamma^r_1)f}
    {1 + 4fK + ff'K^2}.
\end{equation}

\begin{figure}
\centerline{\includegraphics[scale=0.7]{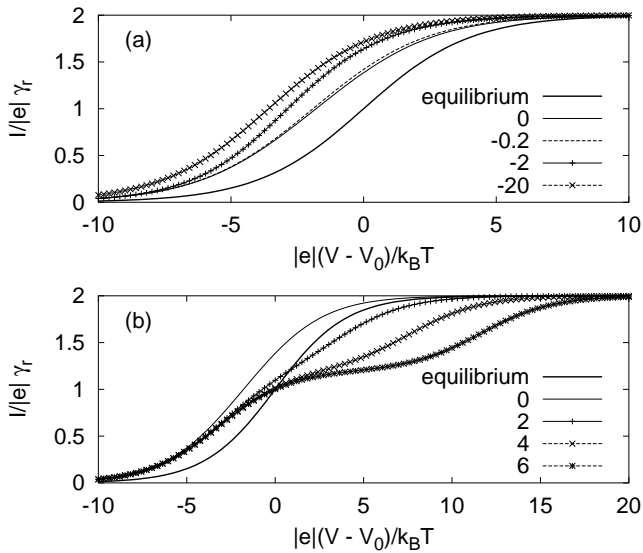}}
\caption{Current steps for different interaction strengths. We assume 
$C_l=C_r$,
$\gamma_1^l=\gamma_2^l=\gamma_l$, and 
$\gamma_1^r=\gamma_2^r=\gamma_r$, with $\gamma_l\gg\gamma_r$. The
``equilibrium'' curve assumes infinitely fast relaxation in the island.
The other curves assume no relaxation and $\delta/k_BT$ ranging (a) from 0
to -20 and (b) 0 to 6.}
\label{interactions}
\end{figure}

\noindent
Fig.~\ref{interactions}(a) shows the current step for the case that
the energy required for the tunneling transition is decreased by
non-equilibrium (negative $\delta$) for various values of
$\delta/k_BT$ ranging from 0 to -20, and Fig.~\ref{interactions}(b)
shows $I$-$V$ curves when the non-equilibrium effect increases the
tunneling energy. These plots were made for $\gamma^r_1 =
\gamma^r_2 = \gamma_r$, $\gamma_l \gg \gamma_r$, and $K=1$. For
negative $\delta$, the effect of the variation in electron-electron
interactions is to produce an additional shift in the
voltage-position of the current step, on top of the shift already
described due to the non-equilibrium states.
This additional shift is proportional to $|\delta|$
if $|\delta|\ll k_BT$ and becomes a constant on the order of $k_BT$
if $|\delta|
\gg k_BT$. A shift of this sort has been observed in Fig.~3(b) of
Ref.~\onlinecite{Mandar01Al}.
For positive $\delta$, the effect of
non-equilibrium is to produce an extra step in the $I$-$V$ curve at
voltages larger than the position of the $\delta=0$ current step.

As $V$ is increased so that more than two levels become
energetically-accessible for tunneling, the ensemble of possible
non-equilibrium excitations grows combinatorically, and each
combination of excitations can produce a different shift for the
tunneling resonance energies. Interactions which depend on the spin
state of the island (neglected thus far) can produce further
complications. The non-equilibrium excitations can produce a variety
of effects depending on the ratio $\gamma_l/\gamma_r$ and on the
magnitude of variations in electron-electron interactions.
When the interaction-induced shifts are comparable to $k_BT$, they have
been observed to produce an effective broadening of the observed 
conductance peaks.\cite{Mandar01Al}
For larger interactions, shifts due to
non-equilibrium excitations have been resolved individually.\cite{Agam97,
Mandar01Co}

\section{Conclusions}

We have solved the rate equations describing electron tunneling via
discrete quantum states on a nanoscale island, for selected simple
cases, under the assumption that rate for internal relaxation of
excited electronic states is slower than the electron tunneling rate.
Even the simplest case of tunneling via a single spin-degenerate
energy level has some initially-surprising features. The magnitude
of the maximum tunneling current can depend on the sign of the
applied bias $V$, and the voltage-position of the resonance is
temperature-dependent. When two spin-degenerate quantum levels are
accessible for tunneling, the behavior is even richer because of the
influence of non-equilibrium excitations on the island. The
voltage-position of the resonance can undergo strong
temperature-dependent shifts even in regimes (\textit{e.g.}, positive 
bias and $\gamma_l
\gg \gamma_r$ noted above) where the one-level resonance positions do
not depend on temperature. Understanding the variations in the
strength of electron-electron interactions is critical in the
non-equilibrium regime with two or more levels accessible. Such
variations can produce additional shifts of resonance curves on top
of the shifts noted previously, and they can also introduce extra
steps into the current-voltage curves.

The methods we have described for determining tunneling currents are
applicable to more than two levels, but the analytic expressions
become sufficiently complicated to be of limited usefulness. We have
verified numerically that the results for additional levels are
qualitatively similar to the two-level case. The computer codes we
have used for calculating the general cases are available
electronically at \url{http://www.ccmr.cornell.edu/~ralph/}.
These are useful, for instance, in extracting the rate-limiting bare
tunneling rates from experimental data in which stepwise increases in
current are measured as $V$ and $V_g$ are adjusted, so that the number
of states accessible for tunneling increases one-by-one.\cite{Mandar01Al}

\section{Acknowledgements}
We thank Piet Brouwer, Abhay Pasupathy, Moshe Schechter, Jan von Delft,
and Xavier Waintal for discussions. This work was supported by the NSF
(DMR-0071631) and the Packard Foundation.

\bibliography{rates}

\begin{thebibliography}{12}
\expandafter\ifx\csname natexlab\endcsname\relax\def\natexlab#1{#1}\fi
\expandafter\ifx\csname bibnamefont\endcsname\relax
  \def\bibnamefont#1{#1}\fi
\expandafter\ifx\csname bibfnamefont\endcsname\relax
  \def\bibfnamefont#1{#1}\fi
\expandafter\ifx\csname citenamefont\endcsname\relax
  \def\citenamefont#1{#1}\fi
\expandafter\ifx\csname url\endcsname\relax
  \def\url#1{\texttt{#1}}\fi
\expandafter\ifx\csname urlprefix\endcsname\relax\def\urlprefix{URL }\fi
\providecommand{\bibinfo}[2]{#2}
\providecommand{\eprint}[2][]{\url{#2}}

\bibitem[{\citenamefont{Ashoori}(1996)}]{Ashoori96}
\bibinfo{author}{\bibfnamefont{R.}~\bibnamefont{Ashoori}},
  \bibinfo{journal}{Nature} \textbf{\bibinfo{volume}{379}},
  \bibinfo{pages}{413} (\bibinfo{year}{1996}).

\bibitem[{\citenamefont{Ralph et~al.}(1995)\citenamefont{Ralph, Black, and
  Tinkham}}]{Ralph95}
\bibinfo{author}{\bibfnamefont{D.~C.} \bibnamefont{Ralph}},
  \bibinfo{author}{\bibfnamefont{C.~T.} \bibnamefont{Black}}, \bibnamefont{and}
  \bibinfo{author}{\bibfnamefont{M.}~\bibnamefont{Tinkham}},
  \bibinfo{journal}{Phys. Rev. Lett.} \textbf{\bibinfo{volume}{74}},
  \bibinfo{pages}{3241} (\bibinfo{year}{1995}).

\bibitem[{\citenamefont{Cobden et~al.}(1998)\citenamefont{Cobden, Bockrath,
  McEuen, Rinzler, and Smalley}}]{Cobden98}
\bibinfo{author}{\bibfnamefont{D.~H.} \bibnamefont{Cobden}},
  \bibinfo{author}{\bibfnamefont{M.}~\bibnamefont{Bockrath}},
  \bibinfo{author}{\bibfnamefont{P.~L.} \bibnamefont{McEuen}},
  \bibinfo{author}{\bibfnamefont{A.~G.} \bibnamefont{Rinzler}},
  \bibnamefont{and} \bibinfo{author}{\bibfnamefont{R.~E.}
  \bibnamefont{Smalley}}, \bibinfo{journal}{Phys. Rev. Lett.}
  \textbf{\bibinfo{volume}{81}}, \bibinfo{pages}{681} (\bibinfo{year}{1998}).

\bibitem[{\citenamefont{Averin et~al.}(1991)\citenamefont{Averin, Korotkov, and
  Likharev}}]{Averin91}
\bibinfo{author}{\bibfnamefont{D.~V.} \bibnamefont{Averin}},
  \bibinfo{author}{\bibfnamefont{A.~N.} \bibnamefont{Korotkov}},
  \bibnamefont{and} \bibinfo{author}{\bibfnamefont{K.~K.}
  \bibnamefont{Likharev}}, \bibinfo{journal}{Phys. Rev. B}
  \textbf{\bibinfo{volume}{44}}, \bibinfo{pages}{6199} (\bibinfo{year}{1991}).

\bibitem[{\citenamefont{Beenakker}(1991)}]{Beenakker91}
\bibinfo{author}{\bibfnamefont{C.~W.~J.} \bibnamefont{Beenakker}},
  \bibinfo{journal}{Phys. Rev. B} \textbf{\bibinfo{volume}{44}},
  \bibinfo{pages}{1646} (\bibinfo{year}{1991}).

\bibitem[{\citenamefont{von Delft and Ralph}(2001)}]{Delft01}
\bibinfo{author}{\bibfnamefont{J.}~\bibnamefont{von Delft}} \bibnamefont{and}
  \bibinfo{author}{\bibfnamefont{D.~C.} \bibnamefont{Ralph}},
  \bibinfo{journal}{Physics Reports} \textbf{\bibinfo{volume}{345}},
  \bibinfo{pages}{61} (\bibinfo{year}{2001}).

\bibitem[{\citenamefont{Agam et~al.}(1997)\citenamefont{Agam, Wingreen,
  Altshuler, Ralph, and Tinkham}}]{Agam97}
\bibinfo{author}{\bibfnamefont{O.}~\bibnamefont{Agam}},
  \bibinfo{author}{\bibfnamefont{N.~S.} \bibnamefont{Wingreen}},
  \bibinfo{author}{\bibfnamefont{B.~L.} \bibnamefont{Altshuler}},
  \bibinfo{author}{\bibfnamefont{D.~C.} \bibnamefont{Ralph}}, \bibnamefont{and}
  \bibinfo{author}{\bibfnamefont{M.}~\bibnamefont{Tinkham}},
  \bibinfo{journal}{Phys. Rev. Lett.} \textbf{\bibinfo{volume}{78}},
  \bibinfo{pages}{1956} (\bibinfo{year}{1997}).

\bibitem[{\citenamefont{Deshmukh
  et~al.}(2001{\natexlab{a}})\citenamefont{Deshmukh, Bonet, Pasupathy, and
  Ralph}}]{Mandar01Al}
\bibinfo{author}{\bibfnamefont{M.~M.} \bibnamefont{Deshmukh}},
  \bibinfo{author}{\bibfnamefont{E.}~\bibnamefont{Bonet}},
  \bibinfo{author}{\bibfnamefont{A.~N.} \bibnamefont{Pasupathy}},
  \bibnamefont{and} \bibinfo{author}{\bibfnamefont{D.~C.} \bibnamefont{Ralph}},
  \bibinfo{journal}{arXiv:cond-mat/0106024}
  (\bibinfo{year}{2001}{\natexlab{a}}).

\bibitem[{\citenamefont{Deshpande et~al.}(1996)\citenamefont{Deshpande,
  Sleight, Reed, Wheeler, and Matyi}}]{Deshpande96}
\bibinfo{author}{\bibfnamefont{M.~R.} \bibnamefont{Deshpande}},
  \bibinfo{author}{\bibfnamefont{J.~W.} \bibnamefont{Sleight}},
  \bibinfo{author}{\bibfnamefont{M.~A.} \bibnamefont{Reed}},
  \bibinfo{author}{\bibfnamefont{R.~G.} \bibnamefont{Wheeler}},
  \bibnamefont{and} \bibinfo{author}{\bibfnamefont{R.~J.} \bibnamefont{Matyi}},
  \bibinfo{journal}{Phys. Rev. Lett.} \textbf{\bibinfo{volume}{76}},
  \bibinfo{pages}{1328} (\bibinfo{year}{1996}).

\bibitem[{\citenamefont{Glazman and Matveev}(1988)}]{Glazman88}
\bibinfo{author}{\bibfnamefont{L.~I.} \bibnamefont{Glazman}} \bibnamefont{and}
  \bibinfo{author}{\bibfnamefont{K.~A.} \bibnamefont{Matveev}},
  \bibinfo{journal}{JETP Lett.} \textbf{\bibinfo{volume}{48}},
  \bibinfo{pages}{445} (\bibinfo{year}{1988}).

\bibitem[{\citenamefont{Deshpande et~al.}(2000)\citenamefont{Deshpande,
  Sleight, Reed, and Wheeler}}]{Deshpande00}
\bibinfo{author}{\bibfnamefont{M.~R.} \bibnamefont{Deshpande}},
  \bibinfo{author}{\bibfnamefont{J.~W.} \bibnamefont{Sleight}},
  \bibinfo{author}{\bibfnamefont{M.~A.} \bibnamefont{Reed}}, \bibnamefont{and}
  \bibinfo{author}{\bibfnamefont{R.~G.} \bibnamefont{Wheeler}},
  \bibinfo{journal}{Phys. Rev. B} \textbf{\bibinfo{volume}{62}},
  \bibinfo{pages}{8240} (\bibinfo{year}{2000}).

\bibitem[{\citenamefont{Deshmukh
  et~al.}(2001{\natexlab{b}})\citenamefont{Deshmukh, Gu\'eron, Bonet,
  Pasupathy, Kleff, von Delft, and Ralph}}]{Mandar01Co}
\bibinfo{author}{\bibfnamefont{M.~M.} \bibnamefont{Deshmukh}},
  \bibinfo{author}{\bibfnamefont{S.}~\bibnamefont{Gu\'eron}},
  \bibinfo{author}{\bibfnamefont{E.}~\bibnamefont{Bonet}},
  \bibinfo{author}{\bibfnamefont{A.~N.} \bibnamefont{Pasupathy}},
  \bibinfo{author}{\bibfnamefont{S.}~\bibnamefont{Kleff}},
  \bibinfo{author}{\bibfnamefont{J.}~\bibnamefont{von Delft}},
  \bibnamefont{and} \bibinfo{author}{\bibfnamefont{D.~C.} \bibnamefont{Ralph}}
  (\bibinfo{year}{2001}{\natexlab{b}}), \bibinfo{note}{submitted to Phys. Rev.
  Lett.}

\end{thebibliography}

\end{document}